hep-ph/9511410
\documentstyle[epsf,12pt]{article}
\newcommand{\bq}{\begin{equation}}
\newcommand{\eq}{\end{equation}}
\newcommand{\ba}{\begin{eqnarray}}
\newcommand{\ea}{\end{eqnarray}}
\newcommand{\nobody}{\rule{0ex}{1ex}}

\newcommand{\ra}{\rightarrow}
\begin{document}
\thispagestyle{empty}
\begin{flushright}
MPI-PhT/95-114\\
LMU-19/95\\
TP-USl/95/06\\
November 1995\\
\vspace*{1.5cm}
\end{flushright}
\begin{center}
{\LARGE\bf Production of $B_c$ mesons in photon--photon and hadron--hadron
collisions\footnote
{Talk presented by K. Ko\l odziej at The Second German-Polish Symposium
{\it ``New Ideas in the Theory of Fundamental Interactions''}, Zakopane,
Poland, 11--15 September, 1995.}
}
\vspace{2cm}\\
Karol Ko\l odziej$^{\rm a,}$\footnote{Supported by the German Federal
Ministry for Research and Technology BMBF under contract
05~6MU93P, and by the EC-program HCM under contract
CHRX-CT93-0132}$^{\rm ,}$\footnote{Supported by the Polish Ministry of
Education under contract No.~PB 659/P03/95/08},
Arnd~Leike$^{\rm b,2}$
        and
Reinhold~R\"uckl$^{\rm b,c,2}$\vspace{0.5cm}\\
$\nobody^{\rm a}${\small\it
 Institute of Physics, University of Silesia,\\
        PL--40007 Katowice, Poland}\\
$\nobody^{\rm b}${\small\it
Sektion Physik der Universit\"at M\"unchen,\\
 Theresienstr. 37, D--80333 M\"unchen, FRG}\\
$\nobody^{\rm c}${\small\it
Max-Planck-Institut f\"ur Physik, Werner-Heisenberg-Institut,\\
F\"ohringer Ring~6, D-80805 M\"unchen, FRG}
\vspace*{2cm}\\
{\bf Abstract}
\end{center}
{\small
We discuss two-photon and hadronic production of $B_c$ mesons in
nonrelativistic
bound state approximation and to lowest order in the coupling constants
$\alpha$ and $\alpha_s$. It is shown that in photon-photon collisions,
heavy quark fragmentation
is dominated by recombination of $\bar b$ and $c$ quarks up to the
highest accessible transverse momenta. In contrast, in
hadroproduction, which at high energies mainly involves
gluon--gluon collisions, the fragmentation mechanism dominates at transverse
momenta $p_T > m_{B_c}$, providing a simple and
satisfactory approximation of the
complete $O(\alpha_s^4)$ results in the high-$p_T$ regime.
Contradictions in previous publications on
hadroproduction of $B_c$ mesons are clarified.
We also present predictions  for cross sections and
differential distributions at present and future accelerators.
}
\vfill\newpage
Since the top quark is too short lived for the formation of quarkonium-like
resonances,
$B_c$ mesons are most probably the only flavoured heavy quark resonances in
nature. Because of flavour conservation in strong and electromagnetic
interactions, the $B_c$ ground state must decay weakly.
The nonrelativistic nature of these bound states provides unique
possibilities to compute genuine nonperturbative quantities such as
fragmentation functions and weak matrix elements, and
to study interesting aspects
of the strong and weak dynamics of hadrons.

A limit on $B_c$ production has been reported recently
by the CDF--collabora\-tion at the Tevatron \cite{cdfbc}. At the LHC, the
production rates are predicted to be large enough for a detailed study of
the production and decay properties \cite{KLR2}.
Also at linear colliders in
the TeV energy range, $B_c$ mesons produced in collisions of
Compton or bremsstrahlung photons may come into experimental reach
\cite{KLR1}.

In this talk, we report about two recent studies \cite{KLR2,KLR1} of $B_c$
production. We discuss
the fragmentation and recombination mechanisms and compare the relative
importance of them in photon--photon and hadron--hadron collisions.
As a main result, we clarify quantitatively the validity of the hard
scattering description in terms of heavy quark fragmentation functions.
Furthermore, on the basis of two completely independent calculations,
we resolve contradictions in previous publications on hadronic
production of $B_c$ mesons \cite{CC}--\cite{MS}.
Finally, we present the most relevant integrated cross sections and
differential distributions as predicted in lowest--order perturbation theory
and nonrelativistic bound state approximation.

In photon--photon collisions, $B_c$ mesons are
produced in association with $b$- and $\bar c$-quark jets:
\begin{equation}
\label{ggbc}
\gamma \gamma \ra B_c b \bar{c}.
\end{equation}
This is also the case in hadronic collisions, where gluon--gluon
scattering,
\begin{equation}
\label{glglbc}
g g \ra B_c b \bar{c},
\end{equation}
is the dominant subprocess at high energies.
In general, one can distinguish two  production mechanisms, namely heavy quark
fragmentation and recombination. We have found these mechanisms to
contribute quite differently in the two reactions, which makes a comparison
very interesting.

The twenty $O(\alpha^2\alpha_s^2)$ Feynman diagrams of process (\ref{ggbc})
can be classified in three gauge invariant subsets characterized
 in Fig.~1: subset (I$_b$) with the $b$-quark
line coupled to the primary photons,
subset (I$_c$) with the $c$-quark line coupled to the primary photons,
and subset (II) with the $b$-quark line coupled to one of the primary
photons and the $c$-quark line to the other one.
The three subsets can be interpreted physically as describing, respectively,
$b\bar b$ production and subsequent $\bar b$ fragmentation,
$\bar b \ra B_c \bar c$, $c\bar c$ production and subsequent $c$ fragmentation,
$c \ra B_c b$, and simultaneous production of a $b\bar b$ and $c\bar c$ pair
and recombination of the $\bar b$- and $c$-quark into a $B_c$ meson.
The gluon--fusion process (\ref{glglbc}) involves a larger number of
Feynman diagrams because of the presence of the gluon self--coupling.
To $O(\alpha_s^4)$, one has thirty--six diagrams in total.
More importantly, they cannot be divided up in gauge invariant subsets
corresponding to the production mechanisms typified in Fig.~1.

In the following, we concentrate on the production of the pseudoscalar
$(B_c)$ and vector $(B_c^*)$ bound states.
Furthermore, we consider the nonrelativistic limit, in which the relative
momentum of the constituents and their binding energy are neglected relative to
the $b$- and $c$-quark masses. In this limit, the masses of the bound
states are equal to the sum of $m_b$ and $m_c$, and the momenta of both
constituent  quarks are proportional to the bound state momentum.
Moreover, the amplitudes for the production of $S$-- waves factorize
into hard scattering amplitudes for $\gamma\gamma$ (or $gg$)
$\rightarrow b\bar b c\bar c$ times the $S$-wave
function at the origin. The latter can be related to the
$B_c^{(*)}$ decay constants. All this results in the
 simple substitution rule \cite {GKPR,KLR1,KLR2}
\begin{equation}
\label{nonrela}
v(p_{\bar b})\bar u(p_c) = \frac{f_{B_c^{(*)}}}{\sqrt{48}}(p\!\!/ - M)
                           \Pi_{SS_Z}\; ,
\end{equation}
indicated in Fig.~1 by the black blob,
where $v(p_{\bar b})$ and $\bar u(p_c)$ denote the $\bar b$- and
$c$- quark spinors, respectively, and
$\Pi_{SS_Z} = \gamma_5 (\rlap/\epsilon)$
is the spin projector for $B_c\ (B_c^*)$.
Note that the colour structure is not accounted for in Eq. (\ref{nonrela}).

\ \vspace{1cm}\\
\begin{minipage}[t]{7.8cm} {
\begin{center}
\hspace{-1.7cm}
\mbox{
\epsfysize=7.0cm
\epsffile[0 0 500 500]{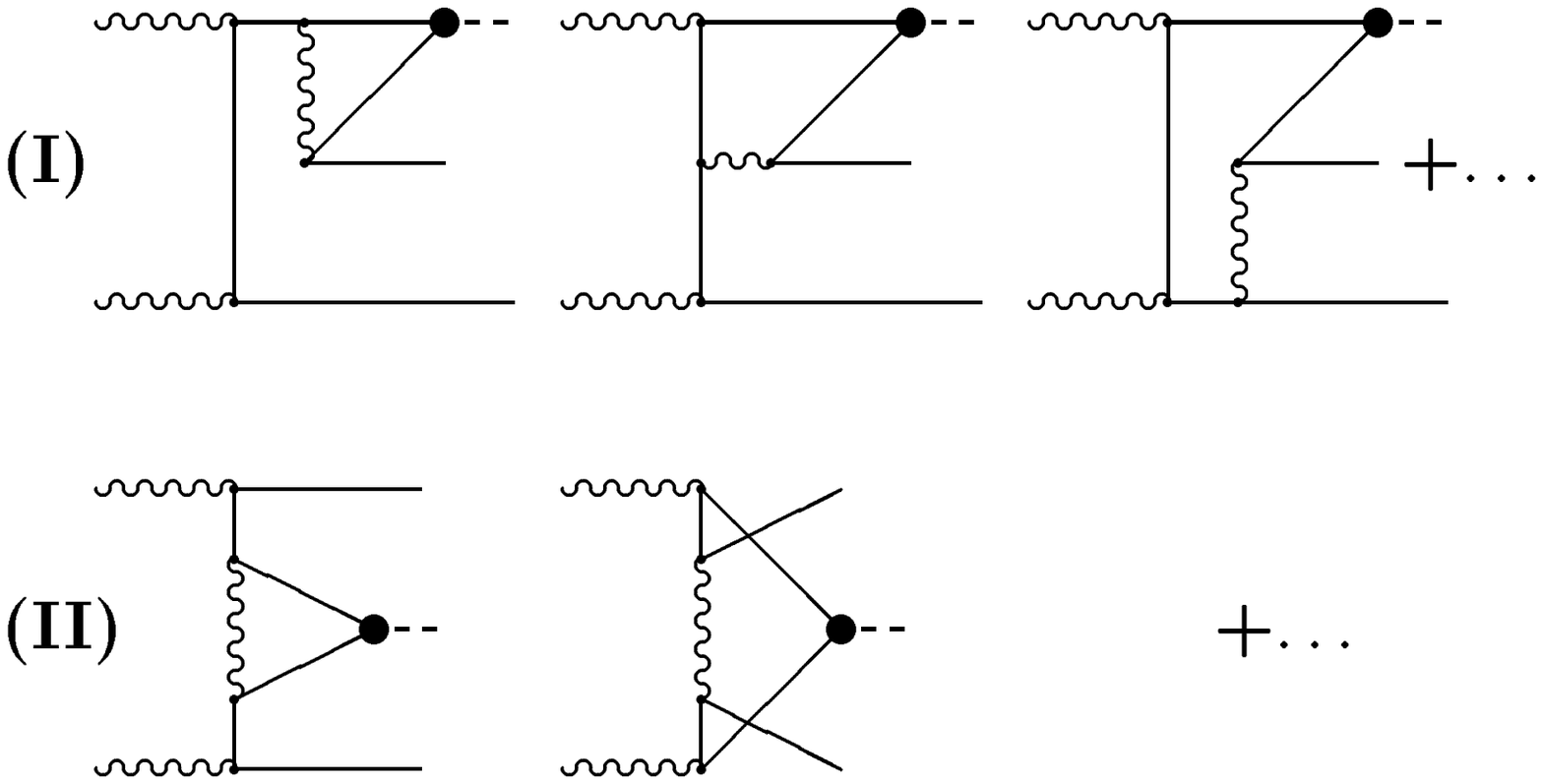}
}
\end{center}
\vspace*{-0.5cm}
\noindent
{\small\bf Fig.~1. }{\small\it
Characteristic topologies of the lowest-order Feynman diagrams contributing
to $\gamma \gamma \rightarrow B_c b \bar{c}$.
}
}\end{minipage}
\hspace{0.5cm}
\begin{minipage}[t]{7.8cm} {
\begin{center}
\hspace{-1.7cm}
\mbox{
\epsfysize=7.0cm
\epsffile[0 0 500 500]{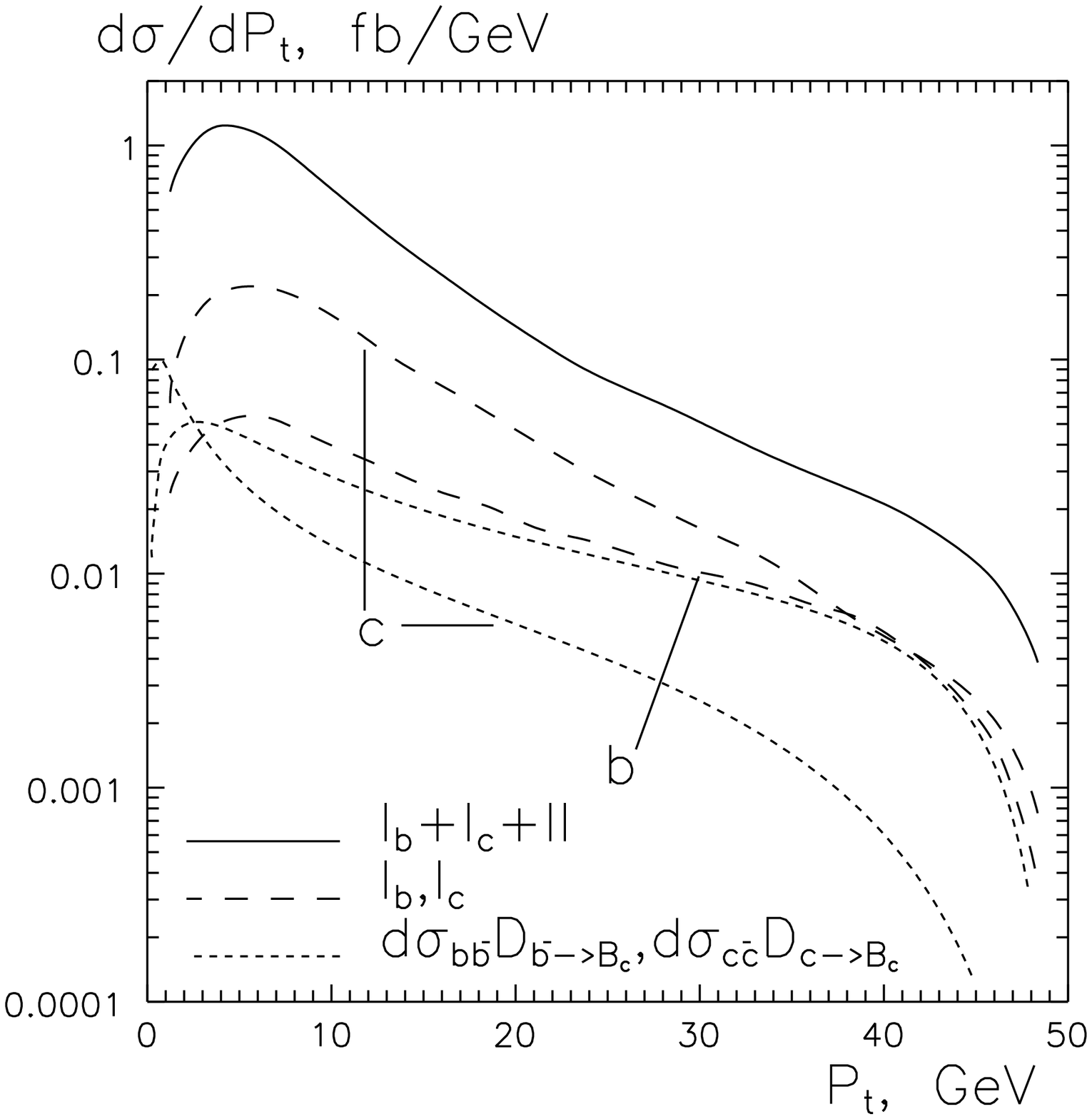}%
}
\end{center}
\noindent
{\small\bf Fig.~2. }{\small\it
Transverse momentum distributions of $B_c$ mesons produced in
$\gamma\gamma \rightarrow B_cb\bar c$ at $\sqrt{s}=100$\ GeV
for the different production mechanisms and approximations
described in the text.
}
}\end{minipage}
\vspace*{0.5cm}

The squared matrix elements for the processes (\ref{ggbc}) and (\ref{glglbc})
have been calculated independently using two different methods,
the traditional
trace technique and the method of helicity amplitudes. The results are found
to be in perfect agreement. In addition, we have checked
external gauge invariance, that is the vanishing of
the matrix elements when the polarization vector of any
of the initial photons or gluons is substituted by its momentum. For process
(\ref{ggbc}), we have also tested internal gauge invariance, that is the
independence
of the matrix element of the gauge parameter in the gluon propagator.
Finally, we have double--checked the phase space integration using two
different Monte Carlo routines.
For more details of the calculation we refer to refs. \cite{KLR2} and
\cite{KLR1}.

In addition to the above $O(\alpha^2\alpha_s^2)$ and $O(\alpha_s^4)$
calculations, we have also studied the factorized description of the
processes (\ref{ggbc}) and (\ref{glglbc}) in terms of
$b\bar b(c\bar c)$ production followed by $\bar b(c)$ fragmentation
\cite{PP2}:
\bq
\label{factappr}
 {\rm d}\sigma_{B_c} = {\rm d}\sigma_{b\bar b}\otimes
D_{\bar b\rightarrow B_c\bar c}(z)\ \ \
({\rm d}\sigma_{c\bar c}\otimes D_{c\rightarrow B_cb}(z)).
\eq
The relevant fragmentation functions $D_{\bar{b}}(z)$ and $D_c(z)$
have been derived from perturbation theory \cite{fragee}. They are known
to provide a perfect approximation of the energy distribution
${\rm d}\sigma/{\rm d}z$ in $e^+e^-\rightarrow B_c\bar c b$
\cite{fragee,teup94} in order $\alpha^2\alpha_s^2$.
Clearly, in photon--photon and hadron--hadron production such a
factorized description cannot be expected to work close to threshold and at
small $p_T$, where the quark  masses play a role.
The question is how well this approximation works at high--$p_T$.
Previous studies of this issue only give qualitative and partly contradicting
answers \cite{CC,BLS}.
This motivated us to compare the factorized approximation (\ref{factappr})
with our complete lowest--order
calculations, and to determine the region of validity of (\ref{factappr})
quantitatively.

The numerical results plotted in Figs.~2-6 have been obtained with
the following values of the parameters:
\begin{eqnarray}
\label{input}
m_b  = 4.8\ {\rm GeV}, \ \ \ m_c  = 1.5\ {\rm GeV},
\ \ \ \alpha = 1/129, \ \ \ f_{B_c} = f_{B_c^*} = 0.4\ {\rm GeV}
\ {\rm\cite{RR}}.
\end{eqnarray}
Furthermore,
we have used the running coupling constant $\alpha_s(Q^2)$
in leading logarithmic approximation for five flavours
and normalized to $\alpha_s(m_Z^2)=0.113$. Additional specifications are
given when needed.

Fig.~2 illustrates transverse momentum distributions of the
$B_c$ in $\gamma\gamma$--production (\ref{ggbc}). We see that
the description (\ref{factappr})
in terms of $b \bar b$ production and
$\bar b$-fragmentation and the corresponding subset (I$_b$) of the diagrams
 of Fig.~1
give distributions (dotted and dashed curves, respectively, labeled by $b$)
which are very similar in shape and normalization
except in the low-$p_T$ region, where the factorized approximation is expected
 to break
down. As far as the shape is concerned this
is also true for $c\bar c$ production and $c$ fragmentation
(curves labeled by $c$).
However, in this case the approximation (\ref{factappr})
fails to reproduce
the correct magnitude of the cross section predicted by the diagrams subset
 (I$_c$) of Fig.~1.
Note that primary $c$--quark production is enhanced to $b$--quark production
by a factor 16 due to the ratio $(Q_c/Q_b)^4$ of the electric charges.
On the other hand, the radiation of a $c \bar c$-pair from a $b$-quark
leads to a harder $p_T$ spectrum for the $B_c$ bound states than
the radiation of a $b \bar b$-pair from $c$-quarks.
Finally, the most important observation
is that the recombination mechanism, represented by the diagrams (II) of
Fig.~1, dominates $B_c$ production not only at
low $p_T$, as one could have expected, but also in the high-$p_T$ region up
to the kinematical limit. In other words, the familiar description of
high-$p_T$ hadron production in terms of the production and fragmentation of
quarks is inadequate for single $B_c$ production
in $\gamma\gamma$-scattering.

The main features of the transverse momentum distributions of the $B_c$
produced
in the subprocess $gg\rightarrow B_c b\bar c$ are illustrated in Fig.~3.
Here, we see that the fragmentation description (\ref{factappr}) indeed
approaches the $p_T$--distributions resulting from the complete
$O(\alpha_s^4)$ calculation, but only in the tails of the distributions.
In order to demonstrate the effect of the quark masses on the fragmentation
kinematics, we have assumed three different relations
between daughter and parent momenta:
\bq
\label{prescription}
p_T = z\sqrt{\hat s/4 -\mu^2}\sin\theta_b,\ \ p_L=p_T\cot\theta_b,\ \
\mu=M\ ({\rm I}),\ \ m_b\ ({\rm II}),\ \ 0\ ({\rm III}),
\eq
where $\sqrt{\hat s}$ is the gluon--gluon c.m. energy.
Case I obeys the physical phase space boundaries. The choices II and
III have been considered in refs. \cite{PP2} and \cite{bls1}, respectively.
Fig.~3 shows that the mass ambiguities of the fragmentation approach
 increase as $\hat s$ decreases and that they become non--negligible at
$\sqrt{\hat s} \le 40\,$GeV.
Comparing the approximations with the $O(\alpha_s^4)$ results, one observes a
slight preference for choice I or II.
Most interesting, however, is the difference
to $\gamma\gamma\rightarrow B_c b\bar c$ where heavy quark fragmentation
is completely subdominant even at large $p_T$.
Apparently, the presence of the gluon self-coupling and
colour factors has a drastic influence on the relative importance
of the fragmentation and recombination.

\ \vspace{1cm}\\
\begin{minipage}[t]{7.8cm} {
\begin{center}
\hspace{-1.7cm}
\mbox{
\epsfysize=7.0cm
\epsffile[0 0 500 500]{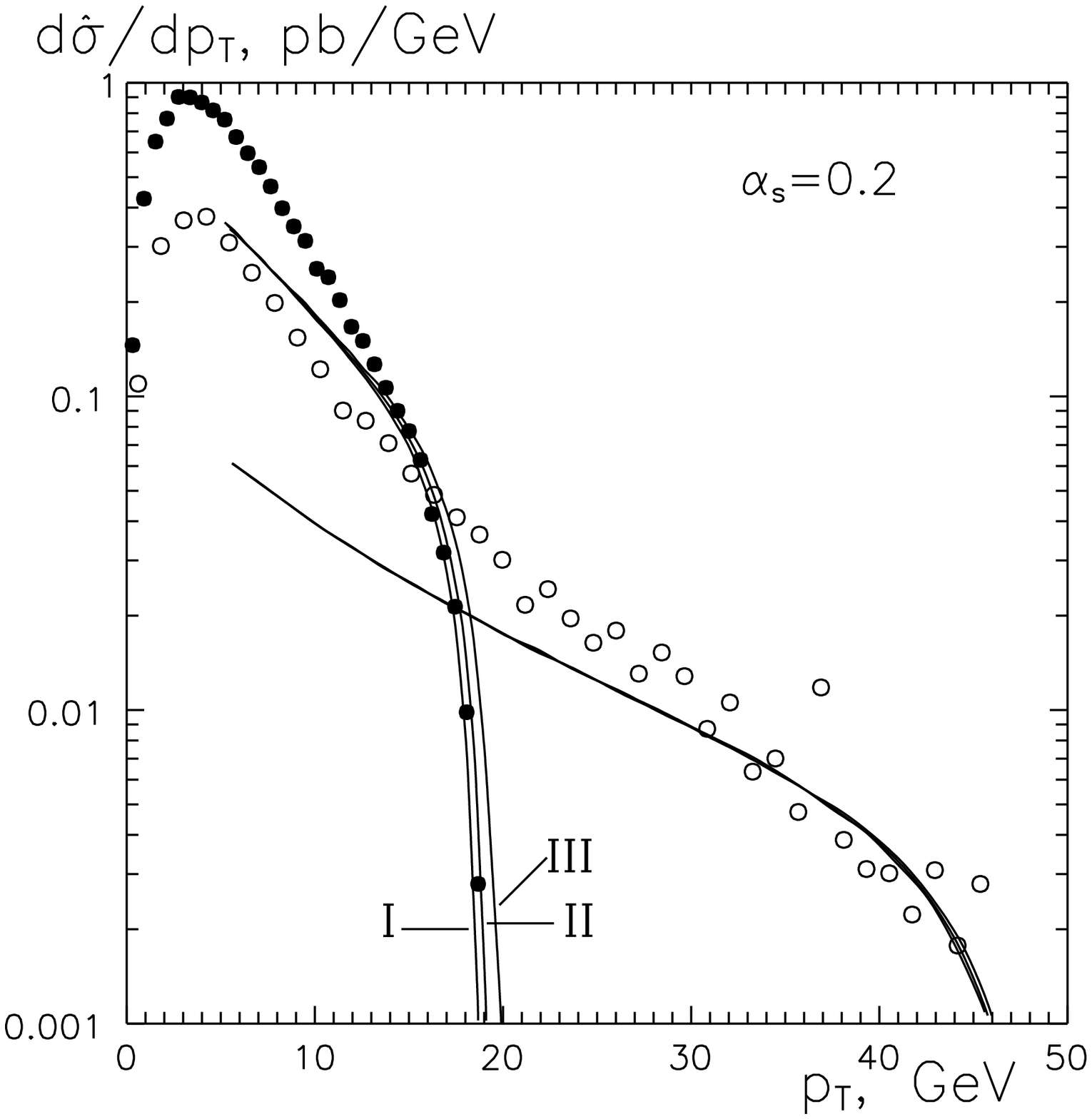}%
}
\end{center}
\noindent
{\small\bf Fig.~3. }{\small\it
Transverse momentum distributions of $B_c$ mesons produced in
$gg\rightarrow B_cb\bar c$ at $\sqrt{\hat s} = 40$ and 100\,GeV:
complete $O(\alpha_s^4)$ calculation (circles) and approximation eq. (4)
(solid curves). The labels {\rm I--III} refer to the kinematics
specified in eq. (6).}}
\end{minipage}
\hspace*{0.5cm}
\begin{minipage}[t]{7.8cm} {
\begin{center}
\hspace{-1.7cm}
\mbox{
\epsfysize=7.0cm
\epsffile[0 0 500 500]{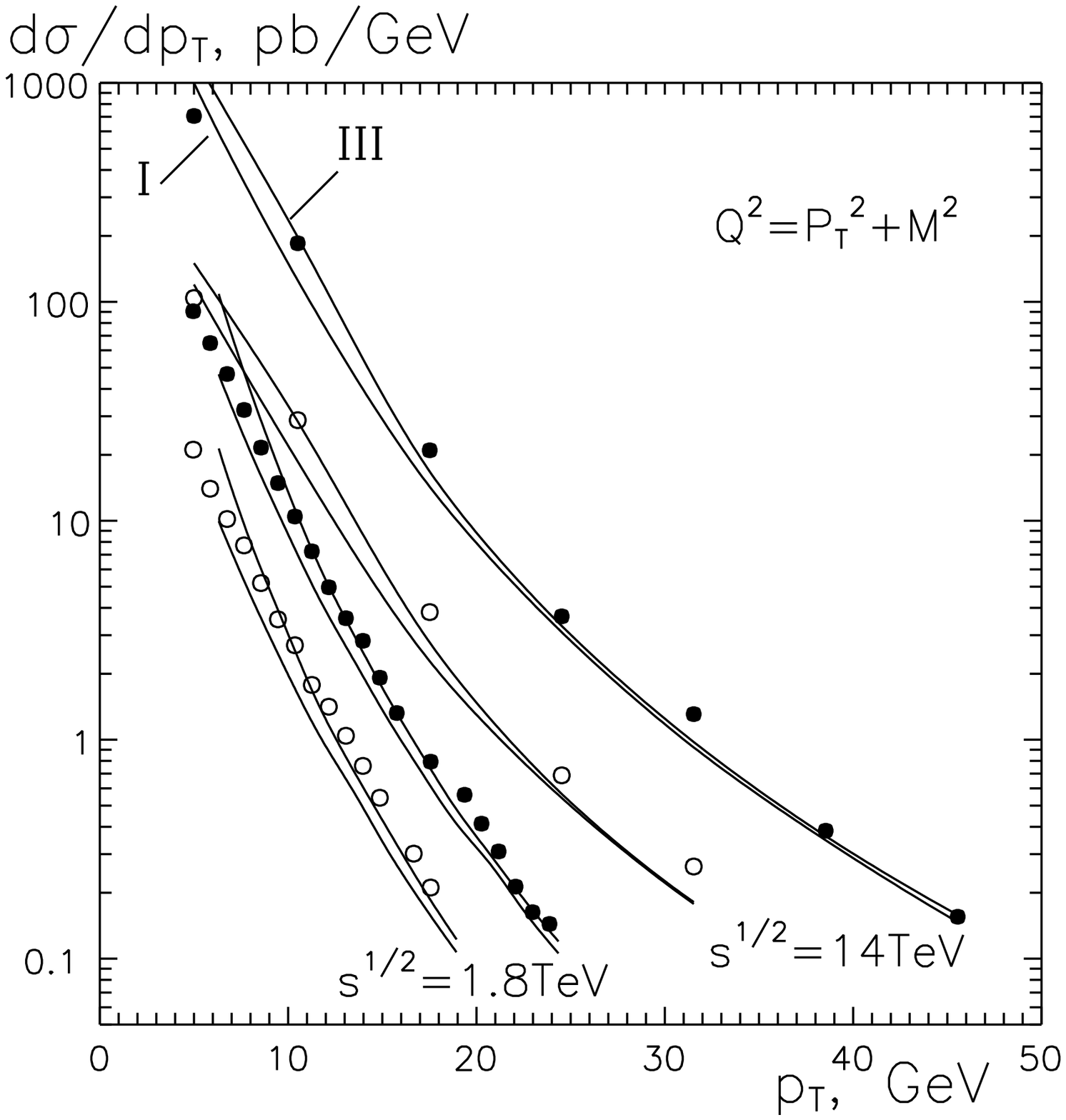}
}
\end{center}
\vspace*{-0.5cm}
\noindent
{\small\bf Fig.~4. }{\small\it
Transverse momentum distribution of the $B_c$
in $p\bar p(pp)$ collisions at the Tevatron (LHC) energies:
complete $O(\alpha_s^4)$ calculation (circles) and fragmentation
approximation eq. (4) (solid curves). The labels {\rm I} and {\rm III}
refer to the kinematics eq. (6).
Results are shown without a rapidity cut (full circles) and for
 $|y|\leq 0.5$ (empty circles).
}
}\end{minipage}
\vspace*{0.5cm}

Predictions for the $p_T$ distributions of $B_c$ mesons in $p\bar p$ and $pp$
collisions at Tevatron and LHC energies are obtained by convoluting
the $gg$-subprocess distributions with the MRS(A$'$) gluon structure
functions \cite{mrs}. The results with and without cuts in rapidity are
plotted in Fig.~4 and  compared with the fragmentation approximation.
We have assumed the scale  $Q^2=p_T^2+M^2$ in both
$\alpha_s(Q^2)$ and the structure functions. The evolution
effects in the fragmentation function $D_{\bar b}(z)$ are ignored
for consistency. These effects are studied in ref. \cite{PP2}.
We see that after convolution the $O(\alpha_s^4)$ calculation and the
fragmentation description (\ref{factappr}) are in reasonable
agreement at $p_T\ge 10\,$GeV.
This can be understood from the properties of the unfolded
$p_T$--distributions illustrated in Fig.~3 and from
the rise of the gluon density at small $x$ which favours contributions from
the smallest possible subenergies $\hat s$, and hence from the tails of the
spectra.
With decreasing $p_T$, the fragmentation picture gradually breaks down and
at $p_T<5\,GeV$ only the complete $O(\alpha_s^4)$ calculation makes sense.
Furthermore, the sensitivity to the kinematical prescription
(\ref{prescription}) decreases slowly with increasing $p_T$.

The total hadronic cross sections for $B_c$ and $B_c^*$ production are shown
 in Fig.~5. The typical rise of  $\sigma$ with energy is due to the rise
of the gluon density as $x$ approaches $x_{min}$ and
the peaking of $\hat\sigma$ near threshold.
In order to demonstrate the scale dependence of $\sigma$,
we have indicated the $B_c$ cross sections at
$\sqrt{s} = 0.1, 1$\ and 10 TeV for different choices of $Q^2$.
As one can see, at lower energies, the notorious scale ambiguity of leading
logarithmic approximations leads to an uncertainty of
more than one order of magnitude. Only at very high energies,
the uncertainty shrinks to a factor two.
It is interesting to note that the differences in the predictions
resulting from different parametrizations of the gluon density, e.g.
CTEQ2 \cite{cteq} instead of MRS(A$'$) \cite{mrs},
would be invisible in Fig.~5.
Other uncertainties, connected with the decay constants $f_{B_c^{(*)}}$
and the effective quark masses $m_b$ and $m_c$, amount at least
to another factor of two.

\ \vspace{1cm}\\
\begin{minipage}[t]{7.8cm} {
\begin{center}
\hspace{-1.7cm}
\mbox{
\epsfysize=7.0cm
\epsffile[0 0 500 500]{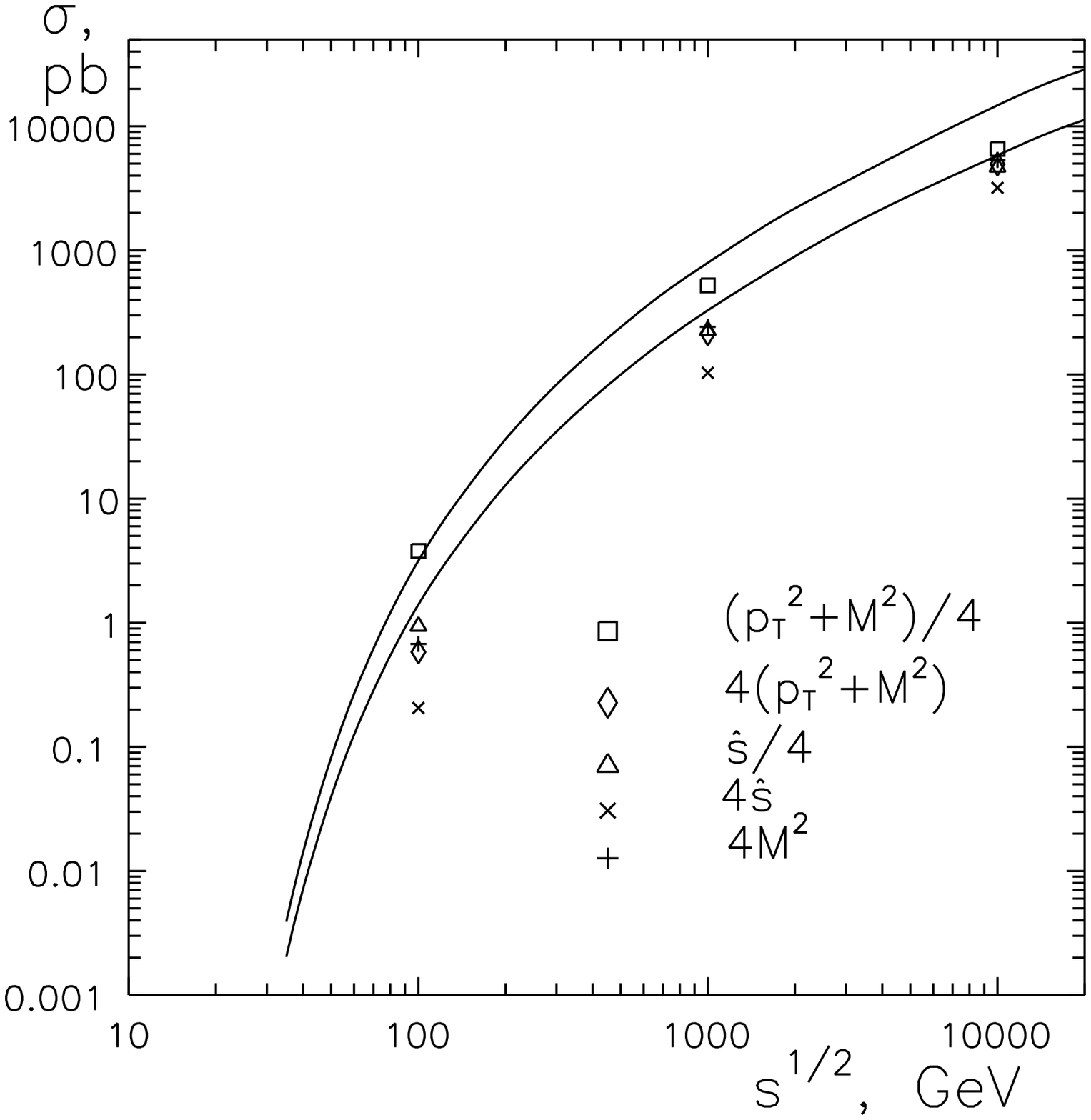}
}
\end{center}
\vspace*{-0.5cm}
\noindent
{\small\bf Fig.~5. }{\small\it
Total cross sections for $pp(p\bar p) \rightarrow B_c^{(*)} b \bar{c} + X$
versus the c.m. energy.
The lower (upper) curve show the predictions for $B_c$ ($B_c^*$)
and the scale $Q^2=p_T^2+M^2$.
The symbols indicate expectations for other choices of $Q^2$.
}
}\end{minipage}
\hspace{0.5cm}
\begin{minipage}[t]{7.8cm} {
\begin{center}
\hspace{-1.7cm}
\mbox{
\epsfysize=7.0cm
\epsffile[0 0 500 500]{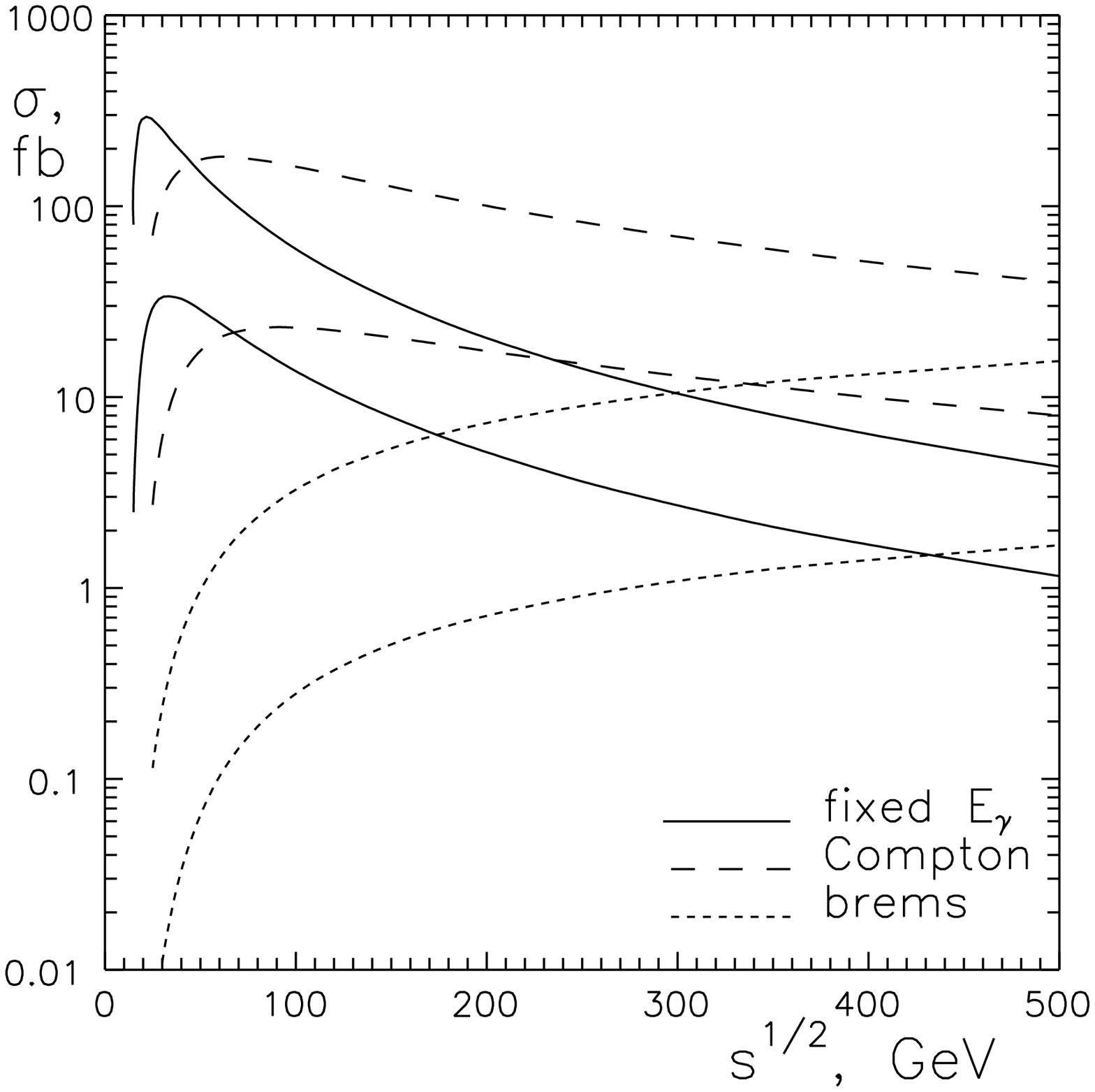}%
}
\end{center}
\noindent
{\small\bf Fig.~6. }{\small\it
Total cross sections for $\gamma\gamma\rightarrow B_c^{(*)}b\bar c$
versus fixed $\gamma \gamma$ c.m. energy,
and after convolution with Compton and bremsstrahlung spectra,
versus the $e^+e^-$ c.m. energy.
The lower (upper) curves correspond to $B_c\ (B_c^*)$ production. The scale of
$\alpha_s$ is chosen to be $p_T^2+M^2$.
}
}\end{minipage}
\vspace*{0.5cm}

The second main motivation for our own studies, besides examining the
validity of the factorized fragmentation approach, has been the wish
to resolve the confusion created by previous calculations \cite{CC} --
\cite{MS} of hadroproduction of $B_c$ mesons which have contradicted each
other. Because of the ambiguities and uncertainties pointed out above,
comparison was not always easy.
We have been very careful in adjusting parameters and the gluon density to
the choice in the respective calculation we considered.
We also focused as much as possible
on the comparison of $gg$ cross sections and distributions
being most transparent. In cases where $gg$ results were not given, we
compared the convoluted cross sections.
The outcome of this comparison was surprising. As described in detail in ref.
\cite{KLR2}, we were not able to reproduce any of the previous numerical
predictions.
In the meantime, the authors of ref. \cite{CC} revised their work \cite{Chen}
and now agree with our results. Also the authors of \cite{BLS} corrected a
normalization error in  \cite{bls1} (this paper appeared almost
simultaneously with \cite{MS}), so that the $gg$ cross section coincide
with what we  found except at very high energies.
However, the $p_T$ distributions given in \cite{bls1}
 still
disagree substantially with ours.

In order to evaluate the observability of $B_c$ mesons at the Tevatron
and LHC, it is useful to integrate the $p_T$--distributions of Fig.~4 over
$p_T\ge p_T^{min}$. Assuming an integrated luminosity of $100\,{\rm pb}^{-1}$,
one can expect about $10^4\ B_c$
mesons with $p_T^{min}= 10\,$GeV at the Tevatron, without taking into
account contributions from the production and decay of $B_c^*$ mesons
and heavier states.
This rate should be sufficient for discovery
in the channel $B_c\rightarrow J/\psi X$, for
which a branching ratio of the order of 10\% is predicted \cite{decay}.
In fact, first results of such a search have already been
reported in \cite{cdfbc}.
Finally, at the LHC for 100${\rm fb}^{-1}$, one can expect $10^7$
 direct $B_c$
mesons  at $p_T^{min}= 20\,$GeV. This rate should then allow a more detailed
study of production and decay properties.

Predictions on the production rates of
$B_c$ mesons in photon-photon collisions at future $e^+e^-$ machines
are obtained by folding the total
cross sections for $\gamma \gamma \ra B_c^{(*)} b \bar{c}$
with the photon spectrum generated by Compton back-scattering of high
intensity laser light
on $e^{\pm}$ beams \cite{Ginzburg}, or with the Weizs\"acker--Williams
bremsstrahlung spectrum \cite{WW}.
The expectations are illustrated in Fig.~6,
where we have plotted the convoluted cross sections for $B_c$ and $B_c^*$
production versus the $e^+e^-$ centre-of-mass energy, together with
the unfolded cross sections as a function of fixed $\gamma \gamma$
centre-of-mass energy.
Because of the long soft tail of the Compton spectrum,
and the shape of the $\gamma\gamma$ cross section
which peaks just above threshold, the convolution
increases the cross sections substantially for energies above 100\,GeV.
At a 500 GeV linear collider and for an integrated luminosity
of 10 fb$^{-1}$, one can produce
 about 100 $B_c$ and 400 $B_c^*$.
The yield of $B_c$ mesons from bremsstrahlung photons is invisibly
small at LEP energies, but increases logarithmically with energy.
In the TeV energy range bremsstrahlung photons become competitive with
back-scattered laser photons in producing $B_c$ mesons.
Although, the prospect of $B_c$ physics in $\gamma\gamma$ collisions are not
very bright, observation of $B_c$ mesons does not appear completely unfeasible.

\end{document}